\documentclass[%
 reprint,
groupedaddress,
 amsmath,amssymb,
 aps,
prc
]{revtex4-1}

\usepackage{graphicx}
\usepackage{dcolumn}
\usepackage{bm}
\allowdisplaybreaks
\usepackage{color}

\newcommand\Peclet{\mbox{\textit{Pe}}}

\usepackage{comment}

\begin{document}

\preprint{APS/123-QED}


\title{Optogenetic switching of migration of contractile cells}

\author{Oliver~M.~Drozdowski}
\author{Falko~Ziebert}%
\author{Ulrich~S.~Schwarz}%
 \email{Corresponding author: schwarz@thphys.uni-heidelberg.de}
\affiliation{Institute for Theoretical Physics, Heidelberg University, Philosophenweg 19, 69120 Heidelberg, Germany}
\affiliation{BioQuant, Heidelberg University, Im Neuenheimer Feld 267, 69120 Heidelberg, Germany}%
\date{\today}

\begin{abstract}
Cell crawling on flat substrates is based on intracellular flows of the actin
cytoskeleton that are driven by both actin polymerization at the front
and myosin contractility at the back. The new experimental tool of optogenetics 
makes it possible to spatially control contraction and thereby possibly
also cell migration.  Here we analyze this situation theoretically
using a one-dimensional active gel model in which the excluded volume interactions of
myosin and their aggregation into minifilaments is modeled by a 
supercritical van der Waals fluid. This physically simple and transparent, 
but nonlinear and thermodynamically rigorous model predicts
bistability between sessile and motile solutions.
We then show that one can switch between these two states at realistic 
parameter ranges via optogenetic activation or inhibition of contractility, in agreement with recent experiments. 
We also predict the required activation strengths and initiation times.
\end{abstract}

\pacs{Valid PACS appear here}
\maketitle

Crawling cell migration is an ubiquitous process in animal tissue and plays a crucial role in
e.g.\ development, wound healing, immune response and cancer metastasis 
\cite{Bodor_2020_DevCell_Cell_shapes_motion_physical_basis_migration}. 
In addition, an increased interest in synthetic cellular systems drives the need for 
understanding the minimal components required for cell motility 
\cite{Goepfrich_2018_TendsBioTech_Bottom_up_construction_synth_cells}. 
The most essential element of cell migration is the symmetry break between
front and back: while the front uses actin polymerization to push
the membrane forward, the back uses myosin II contractility to 
pull the rear forward. Traditionally, these processes are considered
to be coordinated by gradients in biochemical activity, most
prominently the antagonistic signaling pathways of Rac/Cdc42 and RhoA
for front and back, respectively \cite{Ridley_2003_Science_cell_mirgration_integrating_signals}.

A striking feature of locomoting cells is the bistability of their motility behavior:
sessile cells can be stimulated into migration by the application of physical 
stimuli if the applied stimulus is sufficiently large to polarize their cytoskeleton
\cite{Verkhovsky_1999_CurrBiol_Selfpolarization_directional_motility,ziebert_lam}. 
Very recently, it was shown that the direction of cell crawling in channels
can be reversed by optogenetic stimulation effectively decreasing myosin II
contraction at the back \cite{Hadjitheodorou_2021_NatComm_Reorientation_migrating_neutrophils}. 
Because this situation is essentially one-dimensional, 
here the bistability is mediated by polarization, in contrast to
cell migration on two-dimensional substrates, where bistability can
also result from cell shape changes \cite{ziebert_lam,CasademuntPRL}.
Surprisingly, the optogenetic experiments also revealed
that high myosin II activity by itself can prevent reorientation \cite{Hadjitheodorou_2021_NatComm_Reorientation_migrating_neutrophils}.
In general, optogenetic perturbations of cell contractility have
revealed that contractile cells usually do not work at saturation, 
but at an intermediate setpoint of tension that allows for up- and downregulation
\cite{Oakes_2017_NatComm_Optogenetic_RhoA,Valon_2017_Nature_Optogenetic_control_cellular_forces,
Hadjitheodorou_2021_NatComm_Reorientation_migrating_neutrophils}.
These observations shed new light on a long-standing 
question in the fundamental understanding 
of cell migration,  namely how cell migration works
and can be controlled in purely contractile cells.

\begin{figure}[t!]
	\centering
		\includegraphics[width=\columnwidth]{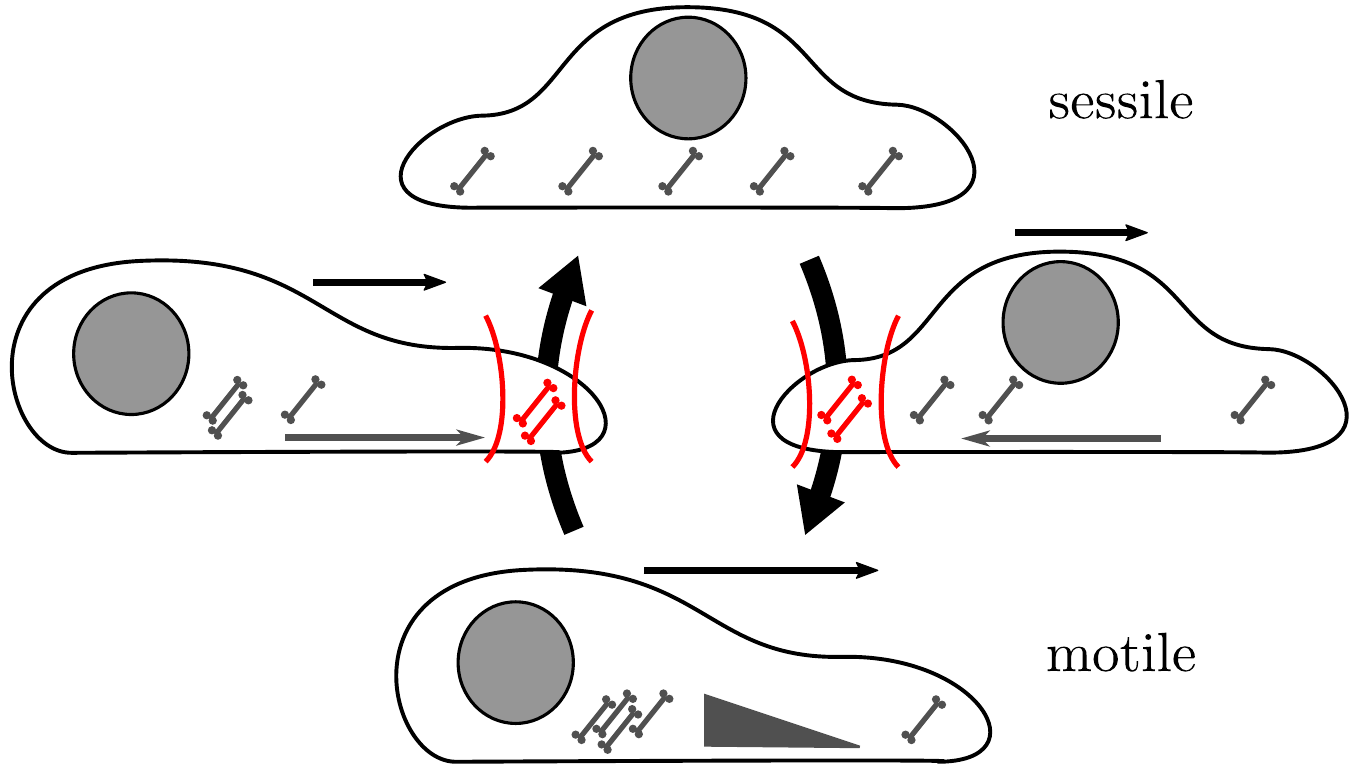}
	\caption{Cell crawling depends strongly on the spatial distribution of myosin motors. A homogeneous distribution does not induce intracellular flows and therefore corresponds to a sessile cell (top). In contrast, a myosin gradient corresponds to a motile cell (bottom).
	Optogenetic activation of contractility (red) can be used to switch between these two states.}
	\label{fig:Schematic_model}
\end{figure}

The natural framework to understand cytoskeletal flow within cells and the role of contraction is active gel theory \cite{KrusePRL04,
Prost_2015_Nature_Review_active_gel_physics}, which has been used 
early on to describe cell migration 
\cite{Kruse_PhysBiol}.
However, obtaining bistability and switching is difficult within this framework. 
Previous attempts have relied on  
introducing nonlinear saturation terms \cite{Recho_2015_JMPS_Mechanics_motility_initiation_arrest, Lavi_2020_PRE_motiltiy_confined_droplet}, but the recent optogenetic experiments \cite{Hadjitheodorou_2021_NatComm_Reorientation_migrating_neutrophils}
demonstrate that the system is not in saturation yet. We also note that 
the earlier work used an ideal gas description for the myosins, which does not reflect
its high density and its property to aggregate into large complexes.

Here we go beyond
these assumptions and propose to describe the myosins as a supercritical van der Waals (vdW) fluid, a concept suggested before for other protein systems 
\cite{Wills_2005_ActCryD_vdw_phase_protein_solutions}.
In the myosin context, it accounts both for the crowdedness of the cytosol and the aggregation of myosin II into so-called
minifilaments, which are supra-molecular clusters that lead to persistent contraction of the actin cytoskeleton. 
The vdW approach results in nonlinearities but is consistent concerning linear irreversible thermodynamics, 
since the driving force, the gradient in chemical potential, is still in linear order. Here we show that this description
provides a physically
reasonable and transparent framework to explain the experimentally observed bistability and to predict the effect of
optogenetic control (Fig.~\ref{fig:Schematic_model}). We also parametrize our model and demonstrate that our 
predictions both agree with the recent 
experiments on motility reorientation \cite{Hadjitheodorou_2021_NatComm_Reorientation_migrating_neutrophils}. 
Finally we use our new theory to predict different experimental protocols to achieve experimental control of cell migration.

To model the cell we assume the constitutive relation of an infinitely compressible active gel with a linear dependence of the active stress on the myosin concentration field $c(x,t)$: 
\begin{equation}\label{eq:constitutive_active_gel}
\eta\partial_x{v}= \sigma - \chi c\ .
\end{equation}
Here $v(x,t)$ is the flow velocity field,
$\eta$ the shear viscosity, $\sigma(x,t)$ the total stress field and $\chi$ the contractility per motor protein.
We assume viscous drag with the substrate, $\partial_x \sigma = \xi v$, with a friction coefficient $\xi$. 
The cell is considered to have a variable length, with left edge $l_-(t)$ and right edge $l_+(t)$, 
and an elastic boundary condition $\sigma=-k(L-L_0)/L_0$, where $L=l_+-l_-$ is the cell length and $L_0$ its reference length. 
We finally assume that the flow fields at the boundaries agree with
cell movement ($v(l_\pm)=\dot{l}_\pm$). This implies that we disregard
the effect of actin polymerization, which could be incorporated as a modified boundary condition \cite{Drozdowski_2021_PRE_Optgen_control_flow_migration}.

To determine the dynamic equation for the myosin concentration, let us start with the
chemical potential $\mu_c$ of the vdW fluid \cite{Johnston_2014_Adv_Thdyn_vdw_fluid}
\begin{multline}
\mu_c = -N_Ak_BT \log \left( \frac{1/N_\mathrm{A}-cb}{cb}\right) +N_A k_BT\ \frac{cb}{1/N_\mathrm{A}-cb} \\ - 2aN_\mathrm{A}^2 c + N_\mathrm{A} k_\mathrm{B} T \log\left(\lambda_\mathrm{th}^{3/2}\right).
\end{multline}
Here $N_\mathrm{A}$ is the Avogadro number, 
$k_\mathrm{B}$ the Boltzmann constant, 
$T$ temperature, 
$b$ the vdW excluded volume 
and $a$ the
average value of the attractive interaction energy per unit concentration. $\lambda_\mathrm{th}=(2\pi\hbar^2/mk_\mathrm{B} T)$
is the thermal wavelength. 
According to linear irreversible thermodynamics \cite{deGroot_1984_Book_non-equilibrium_thermodynamics}, the diffusive particle flux $J_\mathrm{D}$ follows from the
gradient of the chemical potential, 
$J_\mathrm{D} \propto \partial_x \mu $. The continuity equation $\partial_tc = -\partial_x J_\mathrm{D}$ then yields a diffusion equation 
with a concentration-dependent diffusion coefficient
\begin{equation}\label{eq:diffusion_constant_nonliner}
    D(c) = D \mathcal{D}(c)=D \left[ \left(1+\frac{c}{c_\infty-c}\right)^2 - e_\mathrm{A} c\right],
\end{equation}
where we identified the saturation concentration 
$c_\infty = 1/N_\mathrm{A}b$ and the attractive energy 
$e_\mathrm{A} = 2aN_\mathrm{A}/ k_BT$. 
$D(c)$ has a singularity for $c=c_\infty$, reminiscent of 
hard core repulsion, increasing diffusion at high concentrations.
The energetic term leads to a slow down of diffusion for intermediate concentrations due to the attraction.
We limit our discussion to the supercritical vdW fluid, 
i.e., the temperature is above the critical temperature $k_\mathrm{B} T_\mathrm{c} = 8a/27b$ \cite{Johnston_2014_Adv_Thdyn_vdw_fluid}, corresponding
to attractive energies $e_A<e_\mathrm{A}^\mathrm{(c)}=27/4c_\infty$. 
We note that the effects of this
concentration-dependent
diffusion coefficient
have been 
observed experimentally  \cite{Crank_1975_Book_Mathematics_Diffusion},
for instance in the context
of 
binary liquids
\cite{Bardow_2005_JFluid_c_dep_diffusion_ethylacetate_cyclohexane} and
colloidal suspensions of hard spheres \cite{Peppin_2006_JFluidMech_Solidification_colloidal_suspensions}. It has been used in models for bacterial growth \cite{Muller_2002_PRE_Nonlinear_diffusion_bacterial_growth}
and very recently also for excluded volume effects of myosins in cells
\cite{recho_new}.

In addition to performing diffusion, the myosin motors are advected by the active gel, 
$\partial_t c = -\partial_x( v c) + \partial_x( D(c) \partial_x c)$. We consider
no-flux boundary conditions at the edges, i.e., $\partial_xc(l_\pm)=0$.
Following earlier work along these lines \cite{Recho_2015_JMPS_Mechanics_motility_initiation_arrest, Drozdowski_2021_PRE_Optgen_control_flow_migration}, we non-dimensionalize length by $L_0$, time by $L_0^2/D$, stress by $k$, and concentration by $c_0=\int c\,dx / L_0$. We then map the problem on the interval $[0,1]$ using $u=(x-l_-)/L$ and get three dimensionless model parameters: 
the hydrodynamic damping length $\mathcal{L}=\sqrt{\eta/(\xi L_0^2)}$ arising from the competition between viscous and frictional dissipation;
the P\'eclet number $\Peclet=k/\xi D$ describing the importance of advection versus diffusion; and 
myosin contractility $\mathcal{P}=\chi c_0/k$. The inverse P\'eclet number $\mathcal{A}=1/\Peclet$ 
can also be interpreted as adhesion strength, because a large value of $\mathcal{A}$ corresponds to strong friction if $\mathcal{L}^2\mathcal{A}=D\eta/k L_0^2$ is kept fixed, which is a combination of quantities that typically cannot be changed in experiments. 
In the following, adhesiveness $\mathcal{A}$ and contractility
$\mathcal{P}$ are considered as the main parameters, 
as experimentally they are known to control transitions in cell state.

Defining the cell center $G=(l_++l_-)/2$ and the 
advection velocity $\hat{v}(u)=-\dot{G}+\dot{L}(1/2-u)$, and rescaling $\tilde{c}=Lc$,
we finally arrive at our central equations, namely the following boundary value problem (BVP):
\begin{equation}\label{eq:nondim_full_BVP}
\begin{gathered}
\mathcal{L}^2/L^2\ \partial_u^2 \sigma - \sigma = -\mathcal{P}/L\ \tilde{c}, \\
\partial_t \tilde{c} + \frac{1}{L} \partial_u \left[\left(\frac{1}{\mathcal{A}L}\partial_u\sigma + \hat{v}\right) \tilde{c}\right] = \frac{1}{L^2} \partial_u [\mathcal{D}(\tilde{c}/L) \partial_u \tilde{c}],
\end{gathered}
\end{equation}
with the boundary conditions $\sigma(l_\pm) = - (L-1)$ and $\partial_x c(l_\pm) = 0$.
As we will show below, this system can be comprehensively analyzed using a combination of 
analytical and numerical methods.

The parameters of the model can be estimated from experimental data for crawling cells. Following earlier work \cite{Drozdowski_2021_PRE_Optgen_control_flow_migration}, one obtains $\mathcal{L}^2=1.25$ and $\mathcal{P}=0.1$. $\mathcal{A}$ 
can be determined from $k=10^4$\,Pa \cite{Drozdowski_2021_PRE_Optgen_control_flow_migration,Recho_2015_JMPS_Mechanics_motility_initiation_arrest, Loosley_2012, Barnhart_2010}, $\xi = 2\cdot10^{14}$\,Pa\,s\,$/{\rm m}^{2}$ \cite{Drozdowski_2021_PRE_Optgen_control_flow_migration,Barnhart_2011_PLOS_Adhesive_dependent_mechanisms_motile_cell_shape} and $D = 0.7\cdot10^{-12}$\,${\rm m}^{2}/$\,s to be $\mathcal{A}\approx 1/70$. 
A rough estimate for the volume of one (unclustered) myosin motor is $10^2$\,nm$^2$\,$\cdot$\,$100$\,nm and the myosin concentration in cells is of the order of $c_0\approx\mu$M \cite{Robinson_2002_flux_myosin_cytokinesis_concentration}. This implies the estimate $c_\infty=100$, not accounting, however,
for crowding in the cell or finite thickness of the cortex, which 
should decrease this number. Therefore here we 
use $c_\infty=10$, which implies $e_\mathrm{A}^\mathrm{(c)}
=0.675$. 
Note that for the vdW fluid $e_\mathrm{A}=20\pi e_\text{min}/9c_\infty \approx 0.7 \cdot e_\text{min}$ \cite{Johnston_2014_Adv_Thdyn_vdw_fluid},
with $e_\text{min}$ 
(in units of $k_\mathrm{B}T$)
the binding energy in the Lennard-Jones potential of the vdW fluid. 

\begin{figure}[t!]
	\centering
    	\includegraphics[width=\columnwidth]{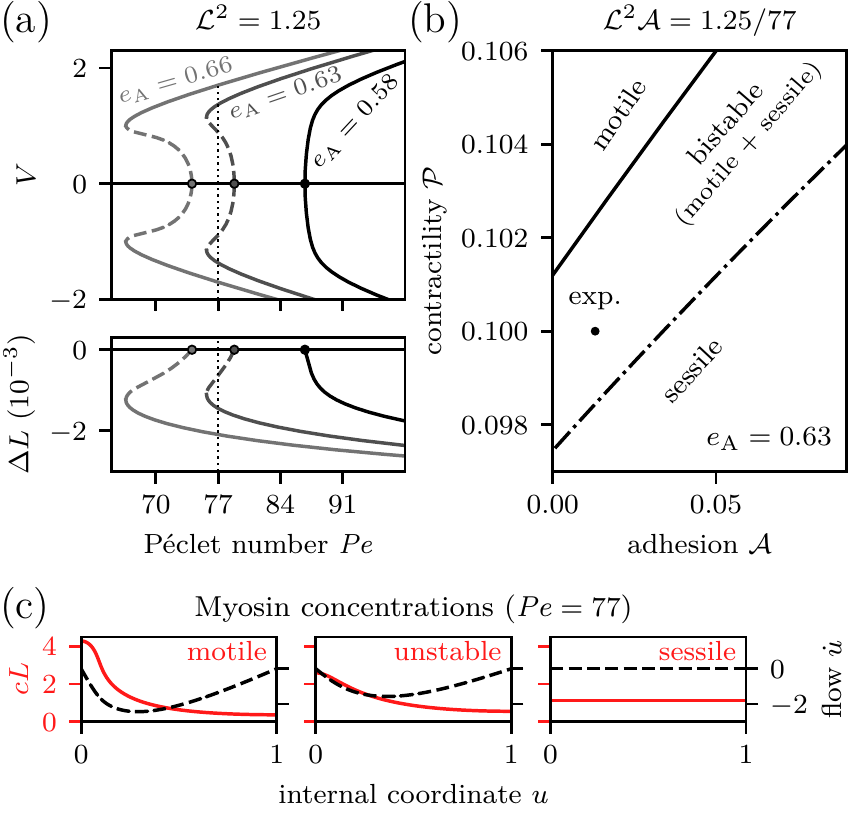}
	\caption{Bistability of cell migration. Panel~(a) gives the cell velocities $V$ and the length differences from the sessile state, $\Delta L = L-\hat{L}_+$, for the obtained
	solution branches as a function of  P\'eclet number $\Peclet$ for different (supercritical) attractive energies $e_\mathrm{A}$. The
	bifurcation points are marked with circles.
	Stable (unstable) solutions are shown as solid (dashed).
	Panel~(b) shows the state diagram for $e_\mathrm{A}=0.63$ and for $\mathcal{L}^2\mathcal{A}=1.25/77$ fixed (containing parameters, which cannot be changed by the cell, cf.~text). Depending on adhesion strength $\mathcal{A}$ and contractility $\mathcal{P}$, one finds a sessile, bistable or motile regime.
	Parameter values estimated from experiments are marked with a circle and used in panel~(c). The solid/dot-dashed curves correspond to the loci of the pitchfork/saddle node bifurcation. Panel~(c) shows normalized motor concentration profiles for experimental parameters (and $V\geq0$) in the bistable regime for the stable motile, the unstable motile and stable sessile solutions,
	as solid curves. The flow velocities are shown in dashed.
	}
	\label{fig:Bifurcation_Diagram}
\end{figure}

We will now study the effect of the nonlinear diffusion and show that it results in bistability between sessile and motile solutions.
To find the steady states one assumes $\dot{l}_\pm=V$, with velocity $V$, $\dot{L}=0$ and steady profiles, i.e., $\sigma$ and $c$ only depend on $u$. One then obtains two coupled ordinary differential equations. 
The case $\mathcal{D}(c)\equiv1$  has already been studied in \cite{Recho_2013_PRL_contraction_driven_motility, Bois_2011_PRL_Pattern_formation_active_fluids}: two non-motile solution families exist with flat stress profiles $\sigma\equiv-(L-1)$ and lengths $\hat{L}_\pm=(1\pm\sqrt{1-4\mathcal{P}})/2$. More complex solutions bifurcate from these branches. However, only two were found to be asymptotically stable, the trivial branch $\hat{L}_+$ and a motile branch bifurcating from it, with a peak in the motor concentration at the trailing edge \cite{Recho_2013_PRL_contraction_driven_motility}. 

For nonlinear diffusion, a flat stress profile is still a steady state solution. In the following we focus on the bifurcation from the sessile, flat-stress state ($L=\hat{L}_+$, 
$\sigma=1-\hat{L}_+$, $\hat{c}=1/\hat{L}_+$) 
to the first motile state, as our numerical results suggest that these are again the only stable solutions for experimentally relevant parameters.
Fig.~\ref{fig:Bifurcation_Diagram}(a) shows the continuation 
in $\Peclet$ of the solutions' cell length and velocity. 
We see that approaching the critical $e_\mathrm{A}^\mathrm{(c)}$ from below renders the supercritical 
pitchfork bifurcation toward the motile solution
to be subcritical, implying bistability. 
Analytically obtained bifurcation points (cf. SI), indicated as circles, agree with the numerics, 
showing that
increasing the attraction $e_A$ in addition decreases the value of $\Peclet$ at which the bifurcation occurs. 
Hence attractive interactions 
both induce bistablity and reduce
the motility threshold.
We also investigated the full, time-dependent BVP numerically,
using the discontinuous Galerkin finite elements method \cite{Ern_2009_JNumAna_discon_galerkin_SWIP_advection_diffusion, Alnaes_2015_FEniCS} (see SI for details),
and found that indeed both solutions
marked in 
Fig.~\ref{fig:Bifurcation_Diagram}(a) 
as solid curves  
are stable in the bistable regime (cf.~SI, Fig.~S1). 

Using advanced continuation methods
(branch point and fold continuation
\cite{Doedel_2007_Book_Lecturenotes_Numerical_Continuation}), we determined the boundaries of the three different regimes
-- sessile, bistable and motile -- 
as shown by the state diagram
in Fig.~\ref{fig:Bifurcation_Diagram}(b). 
Note that we now kept $\mathcal{L}^2\mathcal{A}$ 
fixed as explained above. 
In Fig.~\ref{fig:Bifurcation_Diagram}(b)
we focused on the range around the experimentally reasonable values
$\mathcal{P}\approx 0.1$ and $\mathcal{A} \approx 1/77$ 
(a full diagram can be found in the SI). 
Starting in the bistable regime, 
increasing adhesion leads to a 
transition to the non-motile state, 
while increasing contractility favors
the motile regime. 
Already a change in contractility of $5\%$ allows for these transitions. 
Note that changing $e_\mathrm{A}$ shifts 
all the boundaries
and hence, depending on $e_\mathrm{A}$,
also the opposite scenario can occur, i.e.~decreasing adhesion inducing motility.

Finally, Fig.~\ref{fig:Bifurcation_Diagram}(c) shows the normalized motor concentrations $Lc(u)$ of the three possible states in the 
coexistence region. The 
unstable branch displays 
enrichment of motors 
at the trailing edge, while the motile branch develops a boundary layer.
Importantly, 
the volume exclusion of the vdW 
limits the height of the concentration (and stress) peak at the 
edges even for small $\mathcal{A}$
(cf.~SI Fig.~S1(e)). 
This has to be contrasted 
to the linear model, 
where unrealistically large peaks develop \cite{Recho_2015_JMPS_Mechanics_motility_initiation_arrest}, and is
in agreement with experimental findings of moderate myosin enrichment \cite{Svitkina_Verkhovsky1997}. 
The flow profiles $\dot{u}$ 
shown in Fig.~\ref{fig:Bifurcation_Diagram}(d) indicate the flow to the trailing edge.
The flow velocity is maximum in the
motor-enriched boundary region. 
This,
together with the attraction and subsequent minimum diffusion, 
promotes the formation of the myosin layer.

We stress that for $e_\mathrm{A}=0$
(no attraction, only excluded volume, i.e., the Tonks gas)  
the concentration peaks are still limited, but bistability does not occur.
Thus, accelerating diffusion at high concentrations, as commonly assumed \cite{recho_new, Recho_2013_PRL_contraction_driven_motility}, is not sufficient in this thermodynamically consistent model to achieve bistability. We also find that one needs to be careful using a truncated Taylor expansion as this may result in changes in the criticality of the bifurcation for certain (unphysical) 
parameters of the full excluded volume interaction (see SI for details).

\begin{figure}[t!]
	\centering
		\includegraphics[width=\columnwidth]{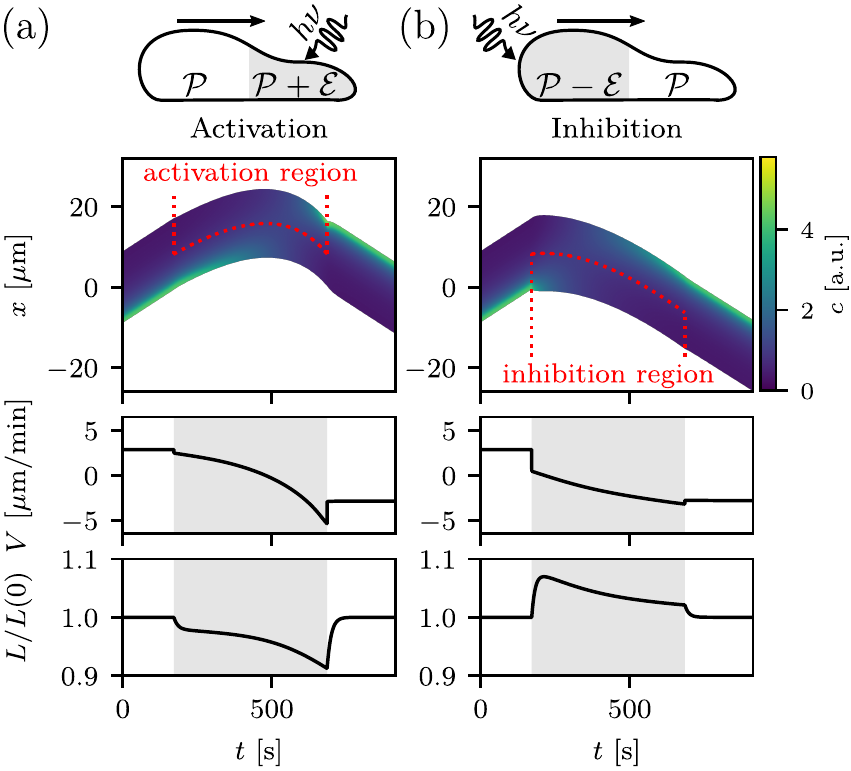}
	\caption{Reversal of cell migration by optogenetics. (a) Positive
	activation strength $\mathcal{E}=0.07$ at the front  and
	(b) negative activation strength $\mathcal{E}=-0.07$ at the back 
	both lead to persistent reversal of migration. 
	The upper panels show kymographs, i.e.~material cell points as function of time in lab coordinates;
	myosin concentration is color-coded.
	The lower panels show cell velocity $V$ and length $L$ scaled by initial length $L(0)$. 
	Time periods of activation/inhibition are shaded. 
	}
	\label{fig:kymograph_act_inhib}
\end{figure}

Having established that the model
displays bistability for experimentally realistic parameters, we next ask if optogenetics
can be used to switch between the sessile and motile solutions.
Optogenetic control of contractility usually exploits light-induced recruitment of a GTP-exchange factor to the cell membrane,
which in turn activates the RhoA signaling pathway and thus leads to an increase in myosin II contractility
\cite{Valon_2017_Nature_Optogenetic_control_cellular_forces, Oakes_2017_NatComm_Optogenetic_RhoA}.
An alternative way 
has been described very recently: there a chemotaxis receptor was optogenetically activated in neutrophils,
which promotes
Rac/Cdc42-activity
and decreases contractility.
Both optogenetic strategies can be implemented in our model by introducing 
an optogenetic contribution to contraction,
$\mathcal{P}\rightarrow \mathcal{P}+\mathcal{E}\Xi$, with a shape function $\Xi$ encoding the spatio-temporal activation
and $\mathcal{E}$ the activation strength
\cite{Drozdowski_2021_PRE_Optgen_control_flow_migration}. 
$\mathcal{E}$ is positive (negative), depending on whether one activates (inhibits) myosin II contractility.
We consider a box-shaped function 
within an activation region $U_\mathrm{act}$, 
i.e., $\Xi(u,t)=1$ only if $u\in U_\mathrm{act}$ and $t\in [t_\mathrm{on},t_\mathrm{off}]$ 
with turn-on/turn-off times
$t_\mathrm{on}$ and $t_\mathrm{off}$.

\begin{figure}[t!]
	\centering
		\includegraphics[width=\columnwidth]{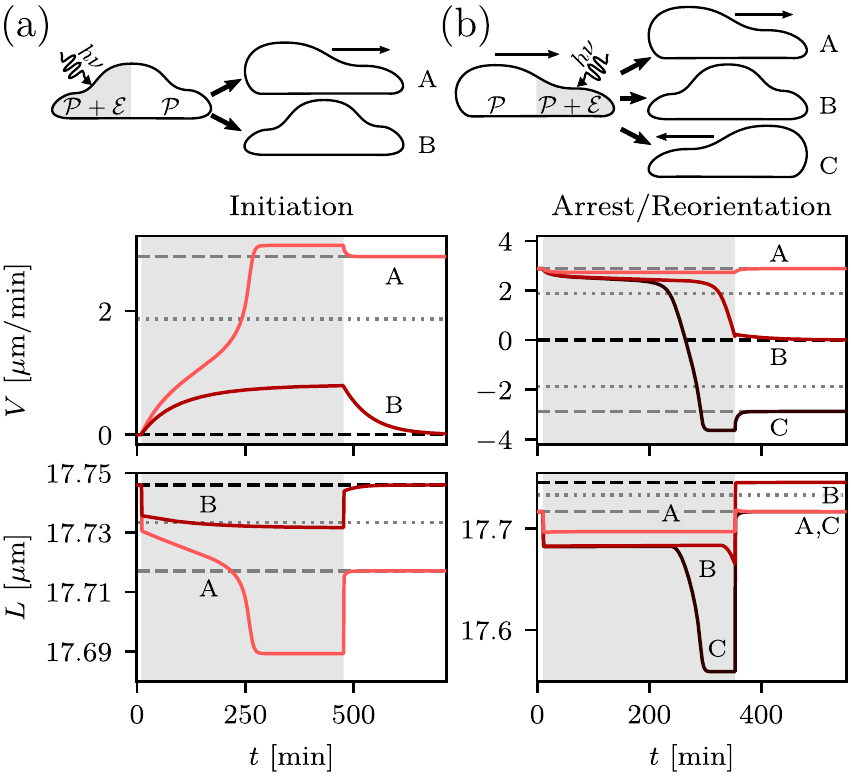}
	\caption{Optogenetic initiation/arrest of motility.
	(a) Motility can be initiated by optogenetically perturbing the flat motor concentration 
	in (the left) half of the cell; activation $\mathcal{E}/\mathcal{P}=1.2\%,  0.8\%$ for A, B. 
	(b) Motility can be arrested or reoriented when a motile steady state is activated in the leading (right) 
	half; activation $\mathcal{E}/\mathcal{P}=6.7\%, 6.6\%, 4\%$ for A, B, and C. Solid curves are simulations and  dashed (dotted)  lines represent
	stable (unstable) steady states; motile (non-motile) states 
	 are  in grey (black).
	Time periods of activation are shaded.}
	\label{fig:optogenetic_perturbation}
\end{figure}

Fig.~\ref{fig:kymograph_act_inhib} shows that 
both activation in the front half (a) and inhibition in the back half (b)
can be used to induce reversals of  direction of cell migration. 
While the first protocol has not been experimentally explored yet, both the simulated trajectories and changes in length for the latter correspond well to the recent experimental results 
\cite{Hadjitheodorou_2021_NatComm_Reorientation_migrating_neutrophils}. Note that initiating motility through activation would also speak in favor of contractility saturation not being central in this context.
In the activation protocol
the cell contracts as contractility is increased. The effect of the perturbation builds up throughout the activation period, since length and velocity are governed by the integrated active stress (see SI). 
For the inhibition scenario
the effect is opposite, as we inhibit in the half with higher initial concentration:
an immediate length response and a more gradual velocity change is obtained.
In particular, we predict a decrease of $|V|$ after turn-off, exactly as
observed experimentally \cite{Hadjitheodorou_2021_NatComm_Reorientation_migrating_neutrophils}.

We next address the question if optogenetic control of contractility can be used to initiate or
arrest motility of cells inside the bistable regime.
In Fig.~\ref{fig:optogenetic_perturbation}(a) we started with
the non-motile steady state
and activated the cell's left half. 
Motility is indeed initiated for $\mathcal{E}$ as small as 1.2\% of $\mathcal{P}$. Increasing $\mathcal{E}$
further leads to faster initiation. 
For the smaller $\mathcal{E}$ shown, 
the perturbation is not sufficiently strong for the induced flow to overcome diffusion and the system cannot leave the basin of attraction of the sessile state.
In Fig.~\ref{fig:optogenetic_perturbation}(b)
we started with
the motile steady state (moving to the right) and activated
the right (leading) half. Compared to the case in (a), now larger perturbations are required, 
because in the motile regime the advection from motility is dominating
and has to be overcome.
Arrest is possible only when fine-tuning
the turn-off of the optogenetic signal: it has to occur in the "re-symmetrized region"
that belongs to the basin of attraction of the sessile solution; 
activating beyond this point rather 
induces reorientation.
Again, larger strengths $\mathcal{E}$ lead to faster arrest/reorientation.

\begin{figure}
	\centering
	\includegraphics[width=\columnwidth]{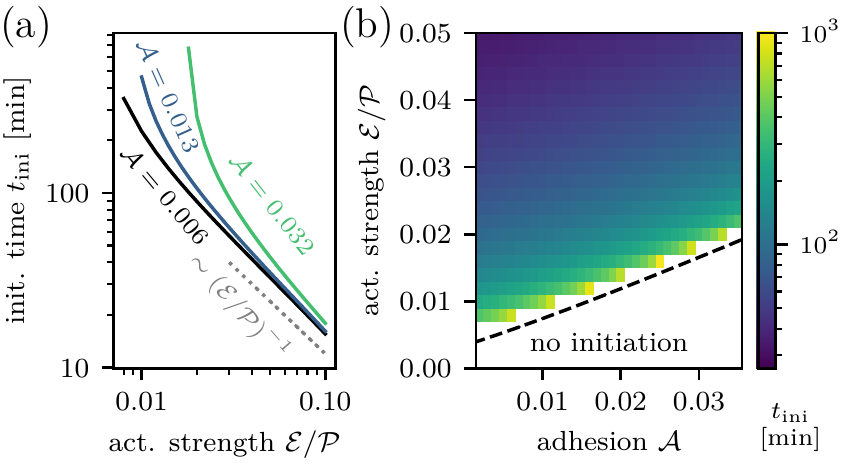}
	\caption{Initiation times of motility in the bistable regime. (a) For different levels of adhesion the initiation time $t_\mathrm{ini}$ for motility differs, where $\mathcal{A}=0.013$ corresponds to $\Peclet=77$. Time $t_\mathrm{ini}$ decreases for larger activations $\mathcal{E}/\mathcal{P}$, asymptotically decaying as $t_\mathrm{ini}\sim(\mathcal{E}/\mathcal{P})^{-1}$. (b) For larger adhesion $t_\mathrm{ini}$ grows larger, with concomitantly increasing minimal activation $\mathcal{E}/\mathcal{P}$ necessary for motility initiation. The dashed line is the initiation threshold, obtained using continuation.}
	\label{fig:optogenetic_initiation_times}
\end{figure}

Having demonstrated the possibility to initiate or arrest cell migration
by optogenetics, we finally predict the corresponding time scales. 
In Fig.~\ref{fig:optogenetic_initiation_times} we show the time $t_\mathrm{ini}$ 
at which the steady state velocity is reached.
We find that $t_\mathrm{ini}$ is larger for stronger adhesion, as one would expect. 
Initiation is faster for larger activation strengths, with an asymptotic dependence of $t_\mathrm{ini}\sim(\mathcal{E}/\mathcal{P})^{-1}$, cf. Fig.~\ref{fig:optogenetic_initiation_times}(a). 
Using continuation of the optogenetically perturbed system (cf. SI), we determined the lower initiation boundary for different adhesions, see Fig.~\ref{fig:optogenetic_initiation_times}(b). 
We find that increasing adhesion not only affects the stability of the steady states, but also slows down the dynamics and increases the necessary activation strength. 
Note that this can be tested experimentally, as activation strength has 
been shown to depend on the illuminated area in optogenetic experiments \cite{Oakes_2017_NatComm_Optogenetic_RhoA}.

This research was conducted within the Max Planck School Matter to Life supported by the German Federal Ministry of Education and Research (BMBF) 
in collaboration with the Max Planck Society. We also acknowledge support  
by the Deutsche Forschungsgemeinschaft (DFG, German Research Foundation) – Projektnummer 390978043.
USS is member of the Interdisciplinary Center for Scientific Computing (IWR) at Heidelberg.

\bibliographystyle{apsrev4-1}
\bibliography{references}{}

\end{document}


\preprint{APS/123-QED}

\title{Supplementary material for: Optogenetic switching of migration of contractile cells}

\author{Oliver~M.~Drozdowski}
\author{Falko~Ziebert}%
\author{Ulrich~S.~Schwarz}%
 \email{Corresponding author: schwarz@thphys.uni-heidelberg.de}
\affiliation{Institute for Theoretical Physics, Heidelberg University, Philosophenweg 19, 69120 Heidelberg, Germany}
\affiliation{BioQuant, Heidelberg University, Im Neuenheimer Feld 267, 69120 Heidelberg, Germany}%
\date{\today}

\pacs{Valid PACS appear here}
\maketitle

\renewcommand{\theequation}{S\arabic{equation}}
\renewcommand{\thefigure}{S\arabic{figure}}

\subsection*{Analytical formula for bifurcation points}
The bifurcation points of steady state solutions of the nonlinear system [Eq.~(4) in the main manuscript] from the trivial branches can be shown to be given by solutions of  
\begin{equation}\label{eq:analytical_bif_loci_implicit_formula}
2 
\left[\cosh(\Omega)-1\right] -
\frac{\mathcal{P} \hat{c}} 
{\mathcal{AD}(\hat{c})}\Omega\sinh(\Omega) = 0,
\end{equation}
with $\Omega^2= [\hat{L}_\pm^2 \mathcal{AD}(\hat{c}) -\mathcal{P}\hat{L}_\pm^2\hat{c}]/[\mathcal{L}^2\mathcal{AD}(\hat{c})]$ and 
$\hat{c}=1/\hat{L}_\pm$. 
This implicit formula, generalizing
a criterion given in
\cite{Recho_2015_JMPS_Mechanics_motility_initiation_arrest} to the nonlinear diffusion
case,
implies that the bifurcation structure 
stays 
similar to the linear case
$\mathcal{D}(c)\equiv1$. 
Figure~\ref{fig:dynamics_relaxation} shows the relaxation dynamics of the model, determining the stability of the branches.
The full bifurcation diagram can be obtained using the continuation method
\cite{Doedel_2007_AUTO07p} and is shown in Fig.~\ref{fig:full_phase_diagram}. 

\subsection*{Description of the used Finite Element Method}
The full, time dependent BVP [Eq.~(4) in the main manuscript] was solved with the discontinuous Galerkin (dG) finite elements method, accounting for the heterogeneous diffusion constant via the symmetric weighted interior penalty scheme for the description of the diffusion term \cite{Ern_2009_JNumAna_discon_galerkin_SWIP_advection_diffusion} and for advection via upwinding. Time was directly discretized and the time-dependent equations for $c$ integrated with an implicit Euler scheme. Length and cell center position were integrated via explicit Euler stepping. The solver was implemented with FEniCS \cite{Alnaes_2015_FEniCS}.

\subsection*{Diffusion of Tonks gas and comparison to Taylor expansion}
For the Tonks gas of hard spheres on a line \cite{Tonks_gas_PR} the nonlinear diffusion coefficient is equivalent to the van der Waals diffusion coefficient without any attractive energy ($e_\mathrm{A}=0$), i.e., analogously to Eq.~(3) in the main text,
\begin{equation}
    \mathcal{D}^\mathrm{Tonks}(c) = \left(1+ \frac{c}{c_\infty-c}\right)^2,
\end{equation}
where $c_\infty = 1/N_Ab$ with hard sphere diameter $b$.
This model, henceforth called the Tonks gas, has the second order Taylor approximation
\begin{equation}
    \mathcal{D}^\mathrm{Taylor}(c) = 1 + \frac{2c}{c_\infty} + \frac{3c^2}{c_\infty^2}.
\end{equation}
Note that the linear term in the van der Waals fluid with attractive energy is $(2/c_\infty-e_\mathrm{A})c$, which means that a diffusion coefficient with vanishing linear term, as suggested in Ref.~\cite{recho_new}, is equivalent to $e_\mathrm{A}=2/c_\infty$, which is still supercritical and thus falls within our framework. Consequently, neglecting a linear term means that attraction has implicitly been considered.

\begin{figure*}[t]
	\centering
		\includegraphics[width=\textwidth]{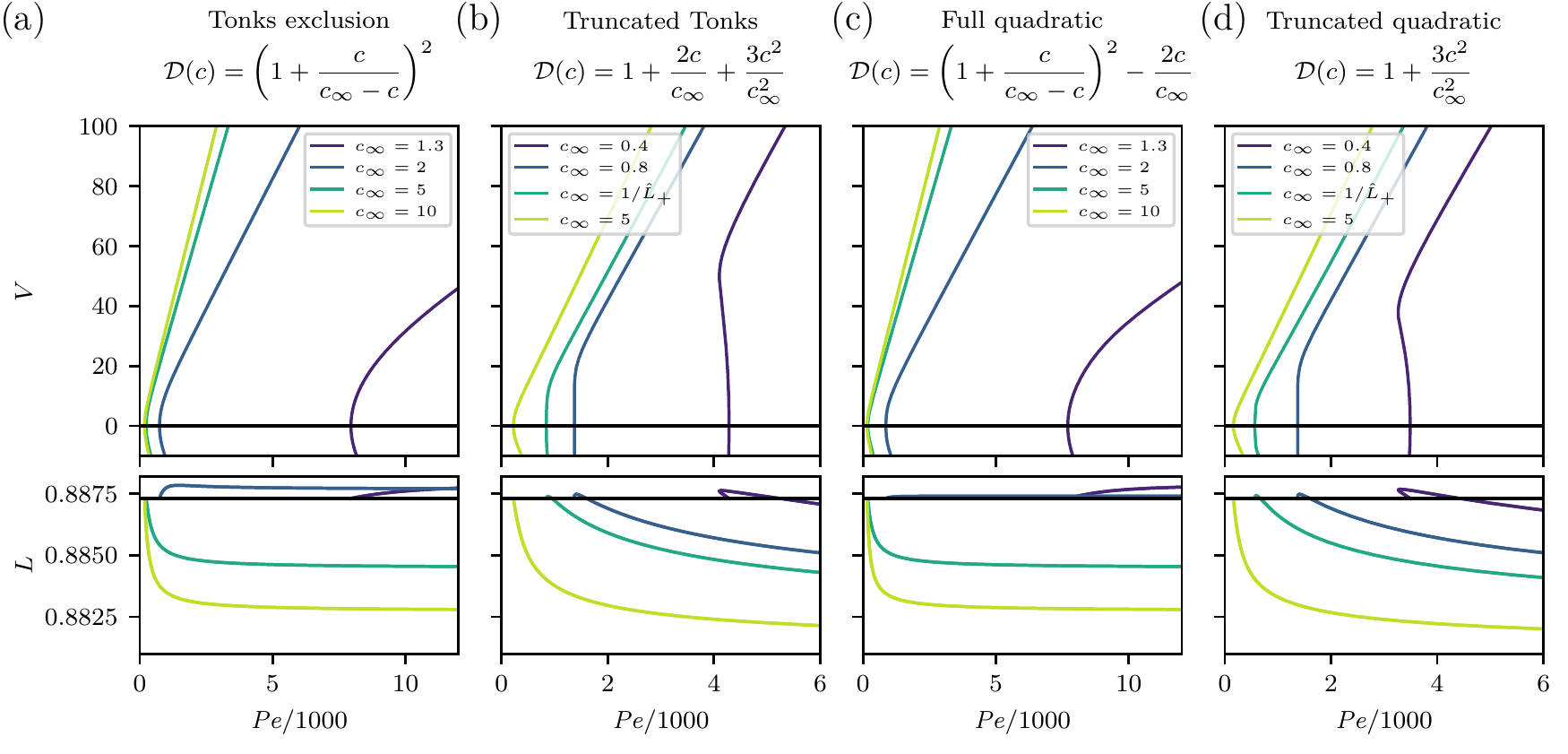}
	\caption{Bifurcation diagrams in nondimensionalized velocity $V$ and length $L$ as functions of the P\'eclet number $\Peclet$ for different nonlinear diffusion models with volume exclusion. (a) Tonks gas model, which is the van der Waals model without attraction. (b) Second order Taylor approximatione of the Tonks gas. (c) Full quadratic model, i.e., the van der Waals fluid with $e_\mathrm{A}=2/c_\infty$, such that the linear term cancels. (d) Second order Taylor approximation of the quadratic model. The term linear in concentration $c$ vanishes.
    In the truncated models we find bistability only in an unphysical regime, see text. Parameters: $\mathcal{P}=0.1$, $\mathcal{L}^2=1.25$}
	\label{fig:bifurcations_diffusion_taylor}
\end{figure*}

\begin{figure*}[t]
	\centering
		\includegraphics[width=\textwidth]{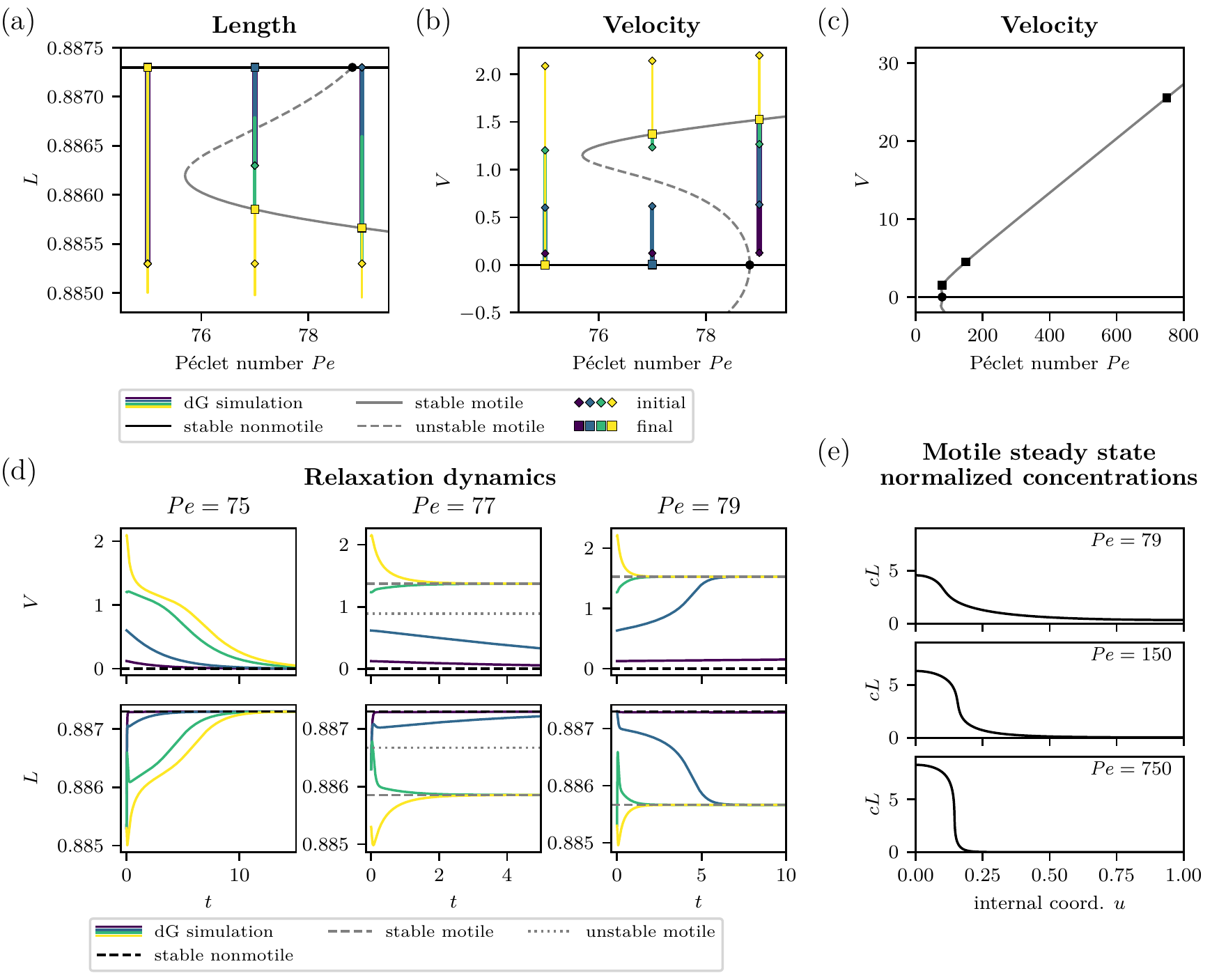}
	\caption{Different initial states relax toward the stable steady states. Panel~(a) shows the bifurcation diagram in length with the trajectories in parameter space of relaxation simulations, shown in Panel~(c). Panels~(b) and (c) show the bifurcation diagrams in velocity in the bistable region and a larger range, respectively. Panel~(d) shows discontinuous Galerkin (dG) simulations of relaxation toward the stable steady states for different P{\'e}clet numbers $\Peclet$.  Depicted are both the velocity $V$ and length $L$ as functions of simulation time $t$. Panel~(e) depicts the normalized concentration profiles $cL$ in internal coordinates $u$ for different $\Peclet$. Note that the relaxation of the length is nontrivial and crossings over multiple solution branches occur. We observe that for large $\Peclet$ a boundary layer develops in the concentration. The different parameters $\Peclet$ are marked in the corresponding bifurcation diagram in Panel~(c) as black squares. Parameters: $\mathcal{P}=0.1$, $\mathcal{L}^2=1.25$, $e_\mathrm{A}=0.63$, $c_\infty=10$.}
	\label{fig:dynamics_relaxation}
\end{figure*}

\begin{figure*}[t]
	\centering
		\includegraphics[width=\textwidth]{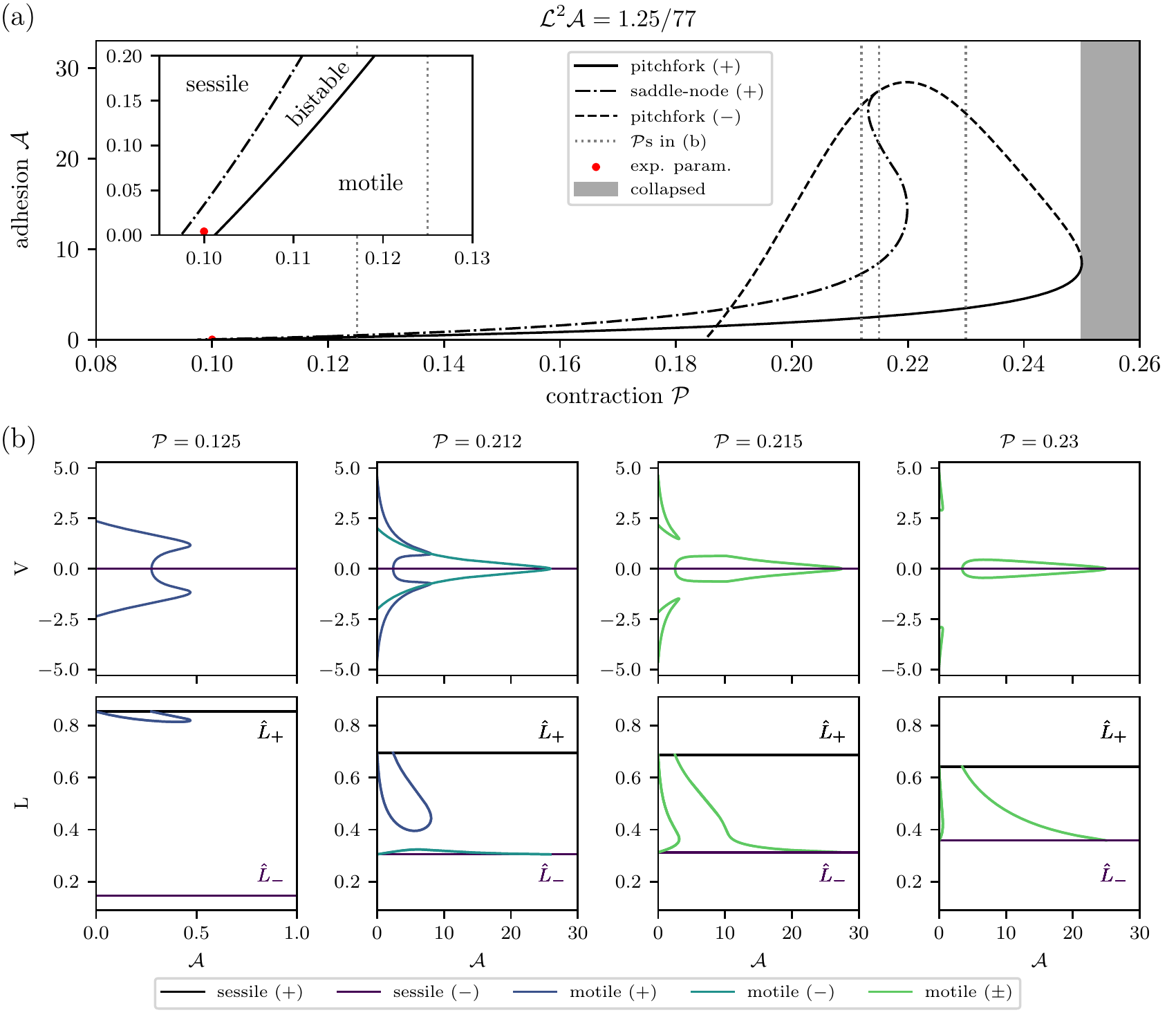}
	\caption{Full diagram showing the loci of bifurcations for our model with $\mathcal{L}^2\mathcal{A}=1.25/77$. For strong contraction $\mathcal{P}\geq1/4$ we have collapse of the solution, as in the linear model. The loci of the pitchfork and saddle node bifurcations have been continued and are shown in Panel~(a). Panel~(b) depicts bifurcation diagrams of velocity $V$ and length $L$ in adhesion $\mathcal{A}$ for different contractilities $\mathcal{P}$, indicated as vertical lines in Panel~(a). Increasing $\mathcal{P}$ from experimental parameters we first obtain a second motile solution bifurcating from the sessile $(-)$ branch, which we have neglected in the main text, as these solutions are not stable or do not exist in the experimentally relevant range. After that the motile branches $(+)$ and $(-)$ merge and we obtain two motile branches, connecting the two sessile $(+)$ and $(-)$ branches, denoted with $(\pm)$. Parameters: $e_\mathrm{A}=0.63$, $c_\infty=10$.}
	\label{fig:full_phase_diagram}
\end{figure*}

Figure~\ref{fig:bifurcations_diffusion_taylor} depicts the bifurcation diagrams for the Tonks gas model, its second order Taylor approximation, the quadratic model with $e_\mathrm{A}=2/c_\infty$ and consistent volume exclusion, and its Taylor approximation, the latter suggested in Ref.~\cite{recho_new}, for different saturation concentrations $c_\infty$. 
Note that in the Taylor approximations we have no divergence at $c=c_\infty$ and thus can look at saturation concentrations below the steady state concentration $c_\infty<\hat{c}=1/\hat{L}_+\approx 1.1$. 
For the Taylor approximation of the Tonks model we find bistability only in this unphysical regime, while including thermodynamic consistent volume exclusion does not yield bistability.
For the truncated quadratic model bistability is generally also in the unphysical region or very close to $c_\infty=1/\hat{L}_+$, not consistent with experiments.

Effectively, truncating the Taylor expansion means that we introduce higher order terms cancelling the higher order terms of volume exclusion. These terms are not consistent with volume exclusion and cannot be neglected as in the nonlinear regime large concentrations are observed \cite{Recho_2013_PRL_contraction_driven_motility}.

\subsection*{Length dynamics governed by integrated active stress}
To see that length and velocity dynamics are governed by the integrated active stress and hence the integrated optogenetic signals, the equation for stress [Eq.~(4) in the main manuscript]
\begin{equation}\label{eq:stress_eq}
\frac{\mathcal{L}^2}{L^2} \partial_u^2 \sigma - \sigma = -\frac{\mathcal{P}+\mathcal{E}\Xi}{L} c,
\end{equation}
is integrated.
The steady state length including optogenetic activation is then given by
\begin{equation}\label{eq:steady_state_length_activation}
\hat{L}_\mathrm{act} = 1+\int_0^1 s(u) \mathrm{d}u - \int_0^1 \left[\mathcal{P}+\mathcal{E}\ \Xi(u)\right]c(u) \mathrm{d}u,
\end{equation}
\vspace{1em}with the stress deviation field from the boundary condition $s(u)=\sigma(u)+(L-1)$. Note that we assumed the signal $\Xi$ to only depend on the internal position $u$, i.e., the cell is in a non-motile steady state or the activation region is moved along with the cell. 
Similarly to Ref. \cite{Drozdowski_2021_PRE_Optgen_control_flow_migration} we find that the steady state length is dominated by the active stress term, i.e., the last integral of the concentration field $c$ in Eq.~(\ref{eq:steady_state_length_activation}).

Using the Green's function for $\sigma$ from Eq.~(\ref{eq:stress_eq}), one can see that the velocity $V$ only depends on the antisymmetric part of the integrated active stress $-(\mathcal{P}+\mathcal{E}\Xi)c$ with respect to the cell center, $u=1/2$, weighted by an appropriate integration kernel from the homogeneous solution. Thus the velocity also depends on the integrated active stress.

\subsection*{Continuation with optogenetic activation}

\begin{figure*}[t]
	\centering
		\includegraphics[width=\columnwidth]{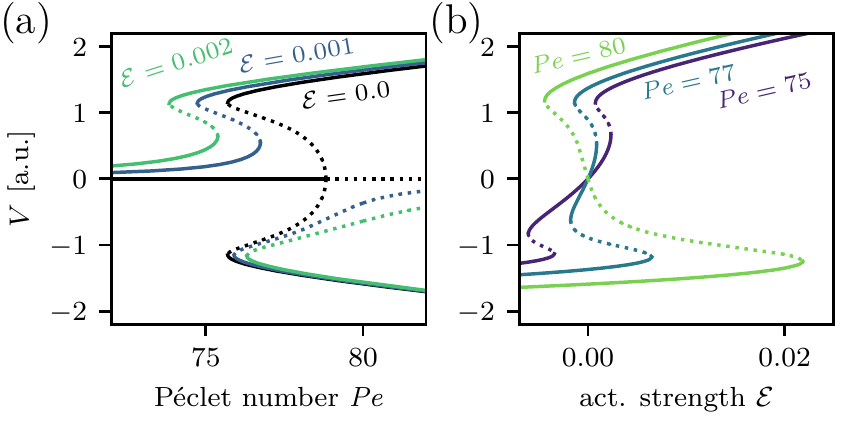}
	\caption{Bifurcation diagram of the model with nonlinear diffusion for optogenetic activation in the left half. (a) The pitchfork bifurcation separates into two saddle-node bifurcations for optogenetic activation with activation strength $\mathcal{E}>0$, which move apart for increasing activation. (b) In the bistable regime we have coexistence of multiple solutions for small activation strengths $\mathcal{E}$ small. However, if the non-motile solution is stable without activation, increasing activation will lead to motility and the cell jumps to the motile solution branch, if activation increases beyond the saddle-node bifurcation. Parameters like in Fig.~2(a) in the main manuscript, i.e., $\mathcal{L}^2=1.25$, $\mathcal{P}=0.1$, $c_\infty=10$, $e_\mathrm{A}=0.63$.}
	\label{fig:continuation_optgen_bifdia}
\end{figure*}

To study the effect of optogenetic activation we included optogenetic activation in the active stress in the equation for stress [Eq.~(4) in the main manuscript], $-\mathcal{P}/L\ c\rightarrow -(\mathcal{P}+\mathcal{E}\Xi)/L\ c$, with a sharp continuous representation of the left-half box function
\begin{equation}
\Xi(u) = \frac{1}{2} \left( 1 + \tanh\left(\frac{1/2-u}{0.001}\right)\right).
\end{equation}
We then used continuation to calculate the bifurcation diagram for different activation strenghts $\mathcal{E}$ for the same parameters as in Fig.~2(a) in the main manuscript with $e_\mathrm{A}=0.63$, see Fig.~\ref{fig:continuation_optgen_bifdia}(a). For optogenetic activation the pitchfork bifurcation separates into two saddle-node bifurcations, where the loss of stability of the previously non-motile solution is marked by the movement of the saddle node-bifuraction beyond the value of $\Peclet$. In Fig.~\ref{fig:continuation_optgen_bifdia}(b) we continue the non-motile solutions ($V=0$) for different $\Peclet$ in $\mathcal{E}$. The saddle-node bifurcation, where branch switching occurs, if activated beyond, marks the activation threshold for motility initiation for fixed $\Peclet$. The stability threshold, shown in Fig.~5 in the main manuscript is determined as the loci of this bifurcation in the $(\mathcal{A},\mathcal{E})$ plane for fixed $\mathcal{L}^2\mathcal{A}=1.25/77$.

\bibliographystyle{apsrev4-1}
\bibliography{references}{}

